\documentclass[10pt,twocolumn]{IEEEtran}

\usepackage{framed}

\usepackage{graphicx}
\usepackage{amssymb}
\usepackage{amstext}
\usepackage{amsmath}
\usepackage{mathtools}
\usepackage{mathrsfs}
\usepackage{times}
\usepackage{array}
\usepackage{multirow}
\usepackage{calc}
\usepackage{array}
\usepackage{subcaption}
\usepackage{definice}
\usepackage{colortbl}
\usepackage{tikz}
\usepackage{pgf}
\usepackage{pgfplots}
\usepackage{enumitem}
\usepackage{booktabs}
\usepackage{pifont}
\usepackage{array}

\usepackage{xcolor}         
\usepackage{mdframed}       
\usepackage{ifthen}         

\def\bfC{\mathbf C}
\def\bfD{\mathbf D}
\def\bfF{\mathbf F}
\def\bfG{\mathbf G}

\def\bfQ{\mathbf Q}
\def\bfP{\mathbf P}

\def\bfS{\mathbf S}

\def\bfU{\mathbf U}
\def\bfV{\mathbf V}

\def\bfa{\mathbf a}
\def\bfb{\mathbf b}

\def\bfg{\mathbf g}
\def\bfh{\mathbf h}

\def\bfu{\mathbf u}
\def\bfv{\mathbf v}
\def\bfw{\mathbf w}
\def\bfx{\mathbf x}
\def\bfy{\mathbf y}
\def\bfz{\mathbf z}

\def\real{\mathbb{R}}

\def\bfxi{{\boldsymbol \xi}}

\def\diag{\operatorname{diag}}

\newtheorem{lemma}{Lemma}

\newcounter{steps}
  {\end{list}}

\def\diag{\mathrm{diag}}

\def\real{\mathbb{R}}

{
  \def\infoboxTopline{true}
  \def\infoboxBottomline{true}
  \ifx#3t\def\infoboxTopline{false}\fi
  \ifx#3b\def\infoboxBottomline{false}\fi
  \mdfsetup{
    frametitle={\colorbox{white}{\space#2\space}},
    frametitleaboveskip=-\ht\strutbox,
    linecolor=#1, 
    linewidth=2pt, 
    roundcorner=10pt, 
    innertopmargin=5pt,
    innerbottommargin=5pt,
    innerleftmargin=10pt,
    innerrightmargin=10pt,
    topline=\infoboxTopline, 
    bottomline=\infoboxBottomline, 
    nobreak=false
  }
  \begin{mdframed}
  \setlength{\parindent}{15pt}
}{
  \end{mdframed}
}

\newcommand{\GrCross}{\raisebox{-0.2mm}{\includegraphics{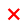}}}
\newcommand{\GrSquare}{\raisebox{-0.2mm}{\includegraphics{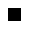}}}
\newcommand{\GrBullet}{\raisebox{-0.3mm}{\includegraphics{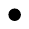}}}
\newcommand{\GrCircle}{\raisebox{-0.3mm}{\includegraphics{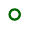}}}

\begin{document}

\thispagestyle{empty}

\begin{center}
\vspace*{3cm}
{\Large\bfseries Preprint Notice}
\vspace{1cm}

This manuscript is a preprint.\\[0.5em]
The final version is currently under review for FUSION 2026. Please cite the published version if accepted.
\end{center}

\clearpage

\setcounter{page}{1}

\title{Tensor Decompositions for Online Grid-Based Terrain-Aided Navigation}
\author{J. Matou\v{s}ek, J. Krej\v{c}\'i, J. Dun\'{i}k, and R. Zanetti
  \thanks{J. Matou\v{s}ek and R. Zanetti are with The University of Texas at Austin, Austin, TX 78712 USA (e-mails: jakub.matousek@austin.utexas.edu, renato@utexas.edu).}%
  \thanks{J. Krej\v{c}\'i and J. Dun\'{i}k are with the Department of Cybernetics, Faculty of Applied Sciences, University of West Bohemia, Pilsen, Czech Republic (e-mails: \{krejci,dunikj\}@kky.zcu.cz).}%
\thanks{The work of J.~Matou\v{s}ek and R.~Zanetti was supported by the 
  Air Force Office of Scientific Research (AFOSR) under award FA9550-23-1-0646. 
  The work of J.~Krej\v{c}\'i and J.~Dun\'{i}k was supported by the Czech Science 
  Foundation (GACR) under grant GA 25-16919J.}%
}
\maketitle

\begin{abstract}
This paper presents a practical and scalable grid\discretionary{-}{-}{-}based state estimation method for high-dimensional models with invertible linear dynamics and with highly non-linear measurements, such as the nearly constant velocity model with measurements of e.g. altitude, bearing, and/or range. Unlike previous tensor decomposition-based approaches, which have largely remained at the proof-of-concept stage, the proposed method delivers an efficient and practical solution by exploiting decomposable model structure—specifically, block-diagonal dynamics and sparsely coupled measurement dimensions.
The algorithm integrates a Lagrangian formulation for the time update and leverages low-rank tensor decompositions to compactly represent and effectively propagate state densities. This enables real-time estimation for models with large state dimension, significantly extending the practical reach of grid-based filters beyond their traditional low-dimensional use.
Although demonstrated in the context of terrain-aided navigation, the method is applicable to a wide range of models with decomposable structure. The computational complexity and estimation accuracy depend on the specific structure of the model. All experiments are fully reproducible, with source code provided alongside this paper (GitHub link: https://github.com/pesslovany/Matlab-LagrangianPMF).
\end{abstract}

\textbf{Keywords:} state estimation, Bayesian inference, nonlinear systems, grid-based filters, Lagrangian filters, CPD, curse of dimensionality, Point-mass filter

\section{Introduction}
State estimation of discrete-time stochastic dynamical systems from noisy measurements has been a subject of significant research interest for decades. Following the Bayesian framework, a general solution to the state estimation problem is provided by the Bayesian recursive relations (BRRs), which compute the probability density functions (PDFs) of the state conditioned on the available measurements. These conditional PDFs offer a complete probabilistic description of the unobservable state of nonlinear or non-Gaussian systems. However, the BRRs are analytically tractable only for a limited class of models, typically those exhibiting linearity. This class of exact Bayesian estimators is represented, for example, by the Kalman filter (KF) for linear and gaussian models \cite{GrAn:15,Si:06}. In all other cases, approximate solutions to the BRRs must be employed. These approximate filtering methods are commonly classified into global and local filters, depending on the validity of their estimates \cite{So:74,SiDu:09}.

Local filters are computationally efficient but can diverge under strong nonlinearity or non-Gaussianity. This paper focuses on global filters, which are more robust but traditionally limited by computational complexity. Two main approaches to solving the BRRs globally employ either stochastic or deterministic numerical integration schemes. Particle filters (PFs), also known as Monte Carlo methods \cite{DoFrGo-book:01}, use stochastic integration, while grid-based filters (GbFs) employ deterministic numerical integration over discretized state spaces.

The baseline GbF, often referred to as the point-mass filter, was introduced in the 1970s \cite{So:74} and later applied to navigation applications \cite{Be:99}. Standard GbFs evaluate conditional PDFs at grid points spanning the continuous state space \cite{SiKraSo:02}, and are generally considered more stable than PFs \cite{AnHa:10,MaDiLiFa:23}. However, their major limitation is the exponential growth in computational and memory requirements with increasing state dimension—a manifestation of the curse of dimensionality—which makes them impractical for problems above four dimensions, even with the state-of-the-art optimized implementations.

This paper proposes a grid-based filter that scales linearly with state dimension, which is made possible by imposing specific structural assumptions on the model. Namely, we assume that the system dynamics matrix is block-diagonal, the dynamics noise covariance matrix is diagonal, so state variables evolve independently (or in loosely\discretionary{-}{-}{-}coupled blocks), and that each measurement dimension depends only on a subset of the state variables. These assumptions hold in many practical applications, such as terrain-aided navigation, or radar and visual tracking, to name a few.

The core of the proposed solution is the canonical polyadic decomposition (CPD) —also known as CANDECOMP or PARAFAC—which represents tensors as sums of rank-one components that are memory efficient compared to the full tensor representation. The entire estimation is performed in this compressed CPD format, i.e., for the rank-one components, and the corresponding full tensors never need to be constructed explicitly. 
The primary advantage of this approach is that it enables scalable filtering in high-dimensional spaces, while fully exploiting model structure for computational savings and improved accuracy. The main drawback is that the CPD rank grows over time, requiring periodic rank reduction (rounding), which introduces approximations and is the main source of error in the method.

The proposed approach is evaluated on real-world data from a terrain-aided navigation (TAN) scenario and is shown to significantly reduce computational complexity even for 4D state estimation, while maintaining high accuracy.

For completeness, we briefly compare the proposed method to other methods that use tensor decompositions to overcome the curse of dimensionality in estimation as well. Namely:
\begin{itemize}
    \item \textbf{Functional decomposition} \cite{TiStDu:23} uses nonnegative tensor factorization to approximate the transient density in a closed region by separating functions of past and future states. While this method was proven effective for mid\discretionary{-}{-}{-}dimensional problems, its scalability appear to be challenging. 
    
    \item \textbf{Tensor-train decomposition} \cite{MaBrDuPu:24} requires both likelihood and transition probability tensors being decomposed to tensor trains at runtime, making it computationally intensive and unstable. Current results remain at the proof-of-concept stage.
    
    \item \textbf{CPD-based methods} \cite{GeUlSpGoKo:24,Go:23} provided the initial inspiration for this work but suffer from several practical limitations. Namely, the grid is fixed (non-moving, non-adapting), the advection step is based on inaccurate finite differences, and model structure is not exploited. 
\end{itemize}

Compared to the mentioned methods that are also based on CPD, the proposed method offers several key advantages:
\begin{itemize}
    \item The grid moves with the state estimate.
    \item The resolution of the grid can adapt during estimation.
    \item The time update (advection) is efficiently handled by the grid motion~\cite{MaDuSt:25}.
    \item Most importantly, the model structure is fully exploited, leading to higher accuracy and significantly lower computational cost.
    \item The estimation is formulated in discrete time, which is the more standard and widely understood approach.
\end{itemize}

The proposed method is implemented using the CPD class and rounding functions from the Tensor Toolbox \cite{tensor_toolbox}. All experiments are fully reproducible using real-world data, and the source code is made available at this link\footnote{https://github.com/pesslovany/Matlab-LagrangianPMF}.

The remainder of the paper is organized as follows. Section~II introduces the principles of grid-based Bayesian estimation. Section~III presents the canonical polyadic decomposition (CPD) and derives the necessary mathematical operations in the CPD format. Section~IV integrates these components to formulate the proposed filter. Section~V verifies the method on a real-world terrain-aided navigation dataset. Finally, Section~VI concludes the paper and outlines directions for future research.

\section{Grid-based Bayesian Estimation}
The model dynamics and measurement equations considered in this paper are, respectively,
\begin{subequations}\label{eq:ass}
\begin{align}
    \bfx_{k+1}& = \bfF_{k}\bfx_{k} + \bfu_k + \bfw_{k}, \label{eq:asx}\\
    \bfz_{k}& = \bfh_k(\bfx_k) + \bfv_k, \label{eq:asz}
\end{align}
\end{subequations}
where $k$ is the time step, $\bfx_k\in\real^{D}$ is the \textit{unknown} state of the system, $\bfu_k$ is the \textit{known} input, and $\bfz_k\in\real^{n_z}$ is the measurement. The matrix $\bfF_k$ is assumed invertible and it describes the state dynamics, and function $\bfh_k$ defines the relation between the state and measurement. State and measurement noises $\bfw_k$ and $\bfv_k$ are unknown, but their PDFs are known. The process noise covariance matrix $\bfQ_k$ is assumed to be diagonal, a limitation the authors intend to address in future work.

\subsection{Bayesian Recursive Relations}
Following the state-space formulation, the filtering task, can be formalised as an estimation of the state $\bfx_k$ based on the available measurements $\bfz^k=[\bfz_0, \bfz_1,\ldots,\bfz_k]$, inputs $\bfu_k$, and the state-space model \eqref{eq:ass}. The \textit{Bayesian} estimation infers the PDF of the state conditioned on available measurements.

The filtering and one-step predictive conditional\footnote{Shorthand notation of the conditional PDF $p(\bfx_{k}|\bfz^{l})=p(\bfx_{k}|\bfz^{l};\bfu^{l-1})$ is used throughout the paper.} PDFs are recursively calculated by the Bayes' rule and the Chapman-Kolmogorov equation (CKE), which form the Bayesian recursive relations (BRRs), as
\begin{align}
    p(\bfx_{k}|\bfz^{k})&\propto p(\bfx_{k}|\bfz^{{k-1}})p(\bfz_{k}|\bfx_{k}),\label{eq:filt}\\
    p(\bfx_{k+1}|\bfz^{k})&=\int p(\bfx_{k+1}|\bfx_{k})p(\bfx_{k}|\bfz^{k})d\bfx_k,\label{eq:pred}
\end{align}
respectively, where $\propto$ denotes equality up to a normalizing constant and where
\begin{itemize}
    \item $p(\bfx_{k}|\bfz^{k})$ is the sought posterior (filtering) PDF at $\mathbf{x}_k$,
    \item $p(\bfx_{k+1}|\bfz^{k})$ is the prior (predictive) PDF at $\bfx_{k+1}$,
    \item $p(\bfz_{k}|\bfx_{k})$ and $p(\bfx_{k+1}|\bfx_{k})$ are the measurement likelihood and state transition PDFs obtained from \eqref{eq:ass}, respectively evaluated at $\bfz_{k}$ and $\bfx_{k+1}$. 
\end{itemize}

\subsection{Point - Mass Density}

\begin{figure}[]
	\centering
	\includegraphics[width=1\linewidth]{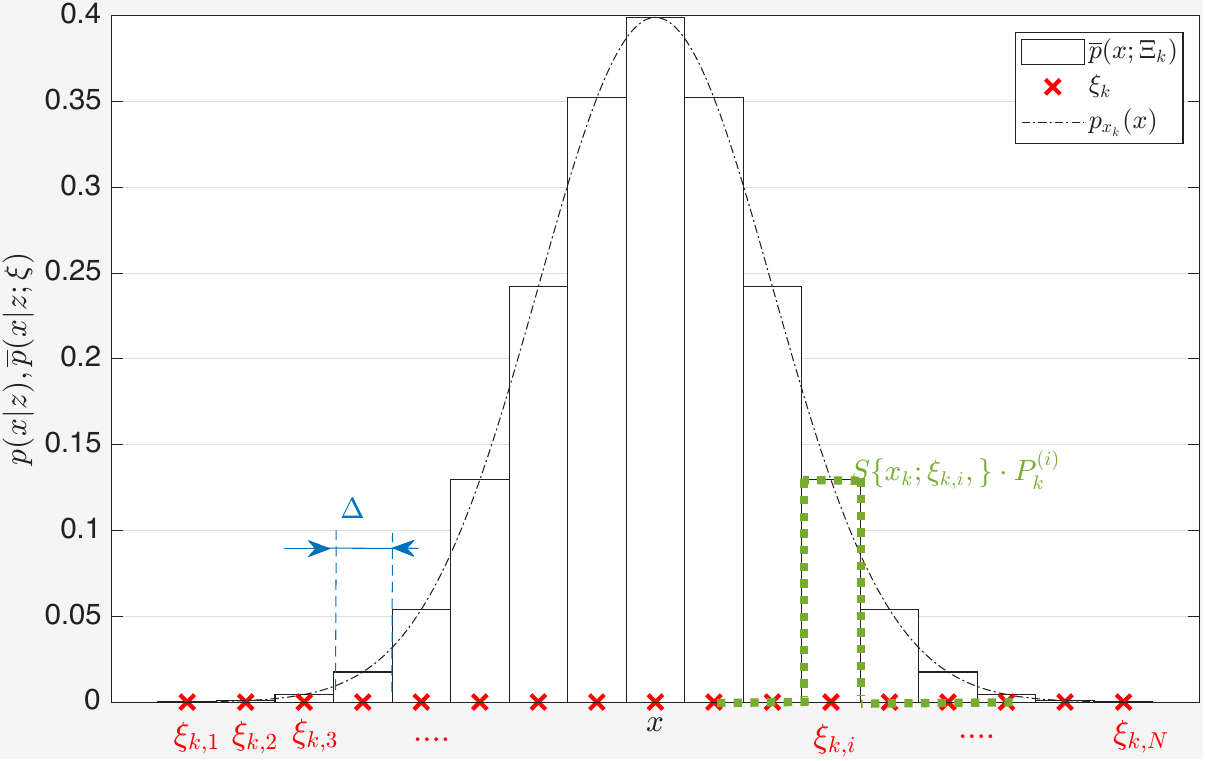}\vspace*{-2mm}
	\caption{Points-mass density}
	\label{fig:PMD}\vspace*{-4mm}
\end{figure}

The grid-based solution to the CKE \eqref{eq:pred} starts with an approximation of the \textit{known} PDF $p_{\bfx_{k}}(\bfx_k)$\footnote{Note that, the lower index RV notation will be dropped for notation simplicity, where possible.} by a \textit{piece-wise constant} point-mass density (PMD) \cite{MaDuBr:23}. The PMD is defined around the set of $N$ grid points $\bfXi_k=\{\bfxi_{k,i}\}_{i=1}^N ,\ \bfxi_{k,i}\in\real^{D}$, as 
\begin{align}
\overline{p}(\bfx_k;\bfXi_k)\triangleq\sum_{i=1}^N{P}_{k}^{(i)}S(\bfx_k;\bfxi_{k,i}),\label{eq:pdf_pm}
\end{align}
where $N = \prod_{j=1}^{D} N_j$, with $N_j$ being a user-defined number\footnote{$N_j$ is usually time independent to achieve constant computational complexity.} of grid points in the $j$-th dimension of the state-space, $P_{k}^{(i)} \propto p(\bfxi_{k,i})$ is a normalised value of the PDF $p(\bfx_k)$ evaluated at the $i$-th grid point $\bfxi_{k,i}$ further called \textit{weight}, and $S(\bfx_k;\bfxi_{k,i})$ is an indicator function that equals to 1 if $\bfx_k$ is in the neighbourhood of the $i$-th point $\bfxi_{k,i}$. That is the weight is constant in the neighbourhood of the $i$-th point. In this paper, an equidistant grid is assumed, i.e., grid where each grid point is associated with the same vector of cell dimension sizes $\bfDelta_{k} \in \real^{D}$ of the same volume $\delta_k$, $ \forall i,$ as illustrated in Figure \ref{fig:PMD}.
The grid boundaries are assumed to be aligned with the state space axes, thus the grid can be represented as a cartesian product 
\begin{align}
	\mathbf{\Xi}_k = {\Xi}_k^1 \times ... \times {\Xi}_k^D \, , \label{eq:axis-aligned-grid}
\end{align}
requiring storage of only $ \sum_{j=1}^D N_j$ values, where $ {\Xi}_k^j \subset \mathbb{R}^{N_j},\ j\in \{1,\dots,D\} $ is a one-dimensional (equidistant) grid.

While the grid was defined as a set, it is implicitly assumed that there is an ordering on the set so that indices of points are readily available at any stage of the algorithm.

In this paper, the weights are stored in a tensor whose order is that of the state space dimension, i.e. $P_k \in \mathbb{R}^{N_1 \times N_2\times \cdots \times N_D}$.
The \emph{modes} of a tensor are further indexed with $i_1, \dots i_D$, leading to $P_k^{(i_1,i_2,...,i_D)}$.
The order within $(i_1,\dots,i_D)$ is often irrelevant, in which case we use a single ``linear''  proxy index $i$ instead, leading to $P_k^{(i)}$, i.e., 
\begin{align}
    i = \underbrace{(i_1,...,i_D)}_{ (\, i_j \, )_{j=1}^D } \in \underbrace{ \{1,\dots,N_1\} \times \cdots \times \{1,\dots,N_D\} }_{ \mathcal{I} } \, . \label{eq:notation-mathcal-I}
\end{align}
With this notation, the normalisation of weights can be conveniently written as $P_k^{(i)} = \frac{P_k^{(i)}}{\delta\sum_{i \in \mathcal{I} } P_k^{(i)}} $. 
Note that a \emph{mode} of a tensor refers to one of its dimensions (axes), i.e., the first mode indexes \emph{rows} while the second \emph{columns}.

\subsection{Lagrangian Grid-based Solution}
The Lagrangian grid-based filter is a state-of-the-art method for efficient state estimation in models with linear dynamics and nonlinear measurements\footnote{Paper on how to extend the Lagrangian approach to any invertible dynamics is under review \cite{DuKrMaBrCh:26}.}, such as~\eqref{eq:ass}, see~\cite{MaDuSt:25}. The approach proposed in this paper builds upon this method.

\subsubsection{Measurement update}The measurement update step, given by \eqref{eq:filt}, applies Bayes’ rule to incorporate the latest measurement information into the conditional PMD, which, in the case of point mass densities, reduces to a simple weight update
\begin{align}
P_{k|k}^{(i)}&\propto  \underbrace{
        p_{\bfv_k}\left(\textbf{z}_k-\bfh_k(\bfxi_{k,i})\right)
    }_{ 
        P_{\mathbf{z}_k | \mathbf{x}_k }^{(i)}
    }
    P_{k|k-1}^{(i)}, \label{eq:measUpdate}
\end{align}
where 
$P_{\mathbf{z}_k | \mathbf{x}_k }^{(i)}$ are the likelihood weights for a fixed $\bfz_k$,
$P_{k|k}^{(i)}$ are the posterior (filtering) weights, and $P_{k|k-1}^{(i)}$ are the prior (prediction) weights, $ \forall i\in \mathcal{I}$~\eqref{eq:notation-mathcal-I}. The measurement update can be given directly for tensors of PMD weights and tensors of likelihood values as
 \begin{align}
	P_{k|k} \propto P_{\bfz_k | \bfx_k } \odot {P}_{k|k-1},\label{eq:measUpdateTen}
\end{align}
where $\odot$ is the Hadamard (element-wise) product. 

\subsubsection{Time-Update}The time-update is performed in two steps. The first step solves the deterministic part of the time-update $\bfx_{k+1}^{\text{adv}} = \bfF_k \bfx_k$ using grid advection (movement)
\begin{align}
	\bfxi_{k+1,i} = \bfF_k \bfxi_{k,i}, \ \forall i \, , \label{eq:flow}
\end{align}
followed by solving the stochastic part of the time-update described by model $\bfx_{k+1} = \bfx_{k+1}^{\text{adv}} + \bfw_k$. For this model the CKE \eqref{eq:pred} is reduced to a convolution (sum of two random variables is a convolution of their PDFs resp. PMDs), i.e. using convolution theorem the PMD evolution is given by
\begin{align}
	P_{k+1|k} = \mathcal{F}^{-1} (\mathcal{F}(P_{k|k}) \odot \mathcal{F}(W_k)), \label{eq:lgbf-convolution}
\end{align}
where $\mathcal{F}^{-1}$ is the inverse Fourier transform and
where $W_k \in \mathbb{R}^{N_1 \times ... \times N_D}$ elements are given by point-wise evaluation of the state noise PDF $p_{\bfw_k}(\bfxi_{k+1}^\mathrm{mean}-\bfxi_{k+1,i})$ with $\bfxi_{k+1}^\mathrm{mean}$ being the sample mean over all points in $\bfXi_{k+1}$.
If $N_j$ is set odd $\forall j$, then $\bfxi_{k+1}^\mathrm{mean}$ is one of the grid points (the ``center'') of $\bfXi_{k+1}$, which further simplifies the calculations.

Several steps (e.g., the interpolation step) were omitted for brevity.
For more information on the standard full tensor approach 
refer to~\cite{MaDuSt:25}. 
In the proposed method, all of the steps are treated in the CPD format that is described next.

\section{Canonical Polyadic Decomposition}
The Canonical Polyadic Decomposition (CPD) \cite{Hi:27} enables the representation of high-order tensors using only a set of factor matrices. Most importantly, the storage and computational complexity of standard operations on tensors in CPD format scale linearly with the number of  state dimension (i.e. order of the tensor), in contrast to the exponential complexity associated with operations on full-format tensors.

The rank-$R$ CPD approximation of a tensor $P \in \mathbb{R}^{N_1 \times \cdots \times N_D}$ is given by a series of outer products as
\begin{align}
{P} \approx \sum_{r=1}^{R} \lambda_P^{(r)} \mathbf{p}_1^{(:,r)} \circ \mathbf{p}_2^{(:,r)} \circ \cdots \circ \mathbf{p}_{D}^{(:,r)}, \label{eq:cpd-general}
\end{align}
where $\circ$ denotes an outer product, $\lambda_P^{(r)}$ is the weight of the $r$-th rank component, and $\mathbf{p}_j^{(:,r)}$ denotes the $r$-th column (loading vector) of the $j$-th factor matrix.
That is, there are $D$ matrices $\bfp_j \in \mathbb{R}^{N_j \times R} , \ j\in\{1,\dots,D\}$. The rank $R$ dictates the accuracy of the CPD representation.

The memory-intensive quantities that must be stored during state estimation are the prior and posterior weights $P$.
The full weight tensors are assumed to be too large to construct explicitly and therefore remain represented by the low-rank elements of the CPD form throughout the entire estimation process.

The proposed algorithm relies on the following operations:
\begin{itemize}
	\item Hadamard (element-wise) product in CPD format.
	\item CPD decomposition of special cases of tensors.
\end{itemize}
These operations are presented in the following subsections.

\subsection{Hadamard Product for the CPD Format}

In both the measurement and time-update steps, a Hadamard (element-wise) product is needed.
Let two tensors $A$ and $B$ have the  CPD
\begin{subequations}    
\begin{align}
A &\approx \sum_{r=1}^{R_A} \lambda^{(r)}_A\, \mathbf{a}_1^{(:,r)} \circ \mathbf{a}_2^{(:,r)} \circ \dots \circ \mathbf{a}_D^{(:,r)}, \\
B &\approx \sum_{s=1}^{R_B} \lambda^{(s)}_B\, \mathbf{b}_1^{(:,s)} \circ \mathbf{b}_2^{(:,s)} \circ \dots \circ \mathbf{b}_D^{(:,s)}.\label{eq:CPDhad}
\end{align}
\end{subequations}
That is, the individual elements of $A$ and $B$ are
\begin{subequations}
    \begin{align}
A^{(i_1, \dots, i_D)} &\approx \sum_{r=1}^{R_A} \lambda^{(r)}_A \prod_{k=1}^D \bfa_k^{(i_k, r)}, \\
B^{(i_1, \dots, i_D)} &\approx \sum_{s=1}^{R_B} \lambda^{(s)}_B \prod_{k=1}^D \bfb_k^{(i_k, s)}.
\end{align}
\end{subequations}

\begin{lemma}
    The Hadamard product $C=A \odot B$ is given by

        \begin{align}
        \label{eq:CPDhadam}
        C \approx & \sum_{r=1}^{R_A} \sum_{s=1}^{R_B} \left( \lambda^{(r)}_A \lambda^{(s)}_B \right)\,
        \left( \mathbf{a}_1^{(:,r)} \odot \mathbf{b}_1^{(:,s)} \right) \circ \cdots \nonumber\\
        &\hspace{3cm} \circ 
        \left( \mathbf{a}_D^{(:,r)} \odot \mathbf{b}_D^{(:,s)} \right),
        \end{align}

    \textit{Proof:}
    Evaluating the product $C=A \odot B$ at the index $(i_1, \dots, i_D)$ reads

        \begin{align}
        C^{(i_1, \dots, i_D)} 
        &\approx \left( \sum_{r=1}^{R_A} \lambda^{(r)}_A \prod_{k=1}^D \bfa_k^{(i_k, r)} \right)
           \left( \sum_{s=1}^{R_B} \lambda^{(s)}_B \prod_{k=1}^D \bfb_k^{(i_k, s)} \right) \nonumber\\
        &= \sum_{r=1}^{R_A} \sum_{s=1}^{R_B} \lambda^{(r)}_A \lambda^{(s)}_B \prod_{k=1}^D \left( \bfa_k^{(i_k, r)}\, \bfb_k^{(i_k, s)} \right), \label{eq:hadCPD}
        \end{align}
  
    whose tensor notation yields~\eqref{eq:CPDhadam}. \hfill $\square$\\
\end{lemma}

It can be seen that the Hadamard product of two CPD tensors results in a new CPD tensor with rank equal to product of their ranks, i.e. $R_A R_B$. The cost of computing each new component involves $D$ Hadamard products of vectors of length $N_j$, costing $\mathcal{O}(\sum_{j=1}^{D} N_j)$ repeated for each pair $(r,s)$ across $R_A \times R_B$ terms.
Hence, the total computational complexity is
\begin{align}
	\mathcal{O}\left( R_A R_B \sum_{i=j}^{D} N_j \right),
\end{align}
This operation thus scales linearly with the tensor order (state dimension) $D$.

\subsection{Special Cases of Decomposition}
During proposed CPD based state estimation routine, certain special cases of tensors need to be decomposed into CPD format. This subsection presents these special cases and their corresponding efficient CPD decompositions.

\subsubsection{Tensor of Initial Condition Weights}

The initial condition random variable $\bfx_0$ is usually normally distributed with a diagonal covariance matrix, thus 
\begin{align}
\bfx_0 \sim 
\mathcal{N}\left(\bfx;\bfmu_0,\diag(\sigma_{0,1}, \dots, \sigma_{0,D})\right) = \ \prod_{j=1}^D \mathcal{N}\left(x^{(j)};\bfmu_0^{(j)},\sigma_{0,j}\right).
\label{eq:ads}
\end{align}
When evaluated at the axes-aligned grid $\bfXi_0$~\eqref{eq:axis-aligned-grid}, the tensor of initial weights $P_{0|-1}$ (for $k=0$) can be written directly in a rank one CPD format as
\begin{align}
P_{0|-1} = P^1 \circ P^2 \circ \dots \circ P^D, \label{eq:decomposition-of-IC}
\end{align}
where $P^j = \bfp_j^{(:,1)}$ is a vector of weights corresponding to the $j$-th Gaussian in~\eqref{eq:ads}.
That is, the tensor~\eqref{eq:decomposition-of-IC} is a special case of the CPD~\eqref{eq:cpd-general}.

If the initial condition is not Gaussian or does not have a diagonal covariance matrix, it can certainly be approximated up to a desired accuracy with a mixture of Gaussians with diagonal covariance matrices; then each single Gaussian will form one rank of the CPD decomposition of $P_{0|-1}$. 

\subsubsection{Tensor with Some Invariant Modes}

A tensor with invariant modes is a tensor whose values do not change along some of its modes, e.g. a matrix whose rows (or columns) are all the same.

Let us consider a tensor $T \in \mathbb{R}^{N_1 \times \cdots \times N_D}$ that varies in only $d < D$ modes. Such a tensor can be fully described by a smaller tensor of unique values, denoted by $M \in \mathbb{R}^{N_1 \times \cdots \times N_d}$. Suppose that $M$ is known, and the goal is to efficiently construct $T$ in CPD format. We focus on three scenarios:
\begin{itemize}
    \item For $d = 1$, $M$ is a vector and can be directly used as a loading vector in the CPD at the $l$-th position, where $l$ is the index of the variant mode. That is,
\begin{align}
T = \sum_{r=1}^{1} \lambda^{(r)} \ \underbrace{ I \circ \cdots \circ I }_{ (l-1)\text{-times} } \circ M^{(:,r)} \circ I \circ \cdots \circ I,
\end{align}
where $I$ is a vector of ones. 
    \item For $d = 2$, $M$ is a matrix. The singular value decomposition (SVD) is a special case of CPD. Therefore, firstly an SVD of $M$ is calculated
\begin{align}
M = \bfU \bfS \bfV^T  = \sum_{r=1}^{R} \bfS^{(r,r)} \mathbf{U}^{(:,r)} \circ \mathbf{V}^{(:,r)} , \label{eq:svd}
\end{align}
and then the CPD is formed by placing the SVD factors at the appropriate positions (given by variant modes indices) as
\begin{align}
T = \sum_{r=1}^{R} \bfS^{(r,r)} \ I \circ \cdots \circ \mathbf{U}^{(:,r)} \circ \cdots \circ \mathbf{V}^{(:,r)} \circ \cdots \circ I. \label{eq:svdForCPD}
\end{align}
The rank $R$ of the SVD decomposition may be truncated based on a user-defined parameter that determines the proportion of the singular values magnitude to be preserved (e.g. 99.99 $\%$).
    \item For $d \geq 3$ there is no shortcut and the tensor $M$ has to be decomposed directly by CPD routine, such as \textit{cp\_als} routine that is part of the Matlab\textcopyright \ tensor toolbox \cite{tensor_toolbox} that was used for the implementation of the proposed method. The rank of the CPD decomposition is set-up based on user defined maximal rank of all CPD decompositions during the estimation. The resulting tensor $T$ can be composed of $M$ analogically as for $d=1$ or $d=2$.
\end{itemize}

\section{High-Dimensional Grid Based Filtering}
This section outlines the proposed method, which employs the weights tensor CPD format representation in the context of high-dimensional nonlinear filtering.
For the proposed method to be efficient, the model~\eqref{eq:ass} must be such that there are subsets of state and measurement variables exhibiting conditional independency.
For convenience, the proposed method is demonstrated on a particular model often utilized in terrain-aided navigation (TAN).
The model comes with a public GitHub repository\footnote{{https://github.com/pesslovany/Matlab-LagrangianPMF}} containing a real dataset and several baseline estimators for comparison, enabling validation of the proposed method’s performance.
The proposed method can be generalized for any model of the form~\eqref{eq:ass} with the said property with little effort.

In the TAN model, it is assumed that the sought state $\bfx_k$ with $D=4$ (i.e. state estimation dimension $D=4$) contains the vehicle \textit{horizontal} position $\bfp_k^\mathrm{W}=[p_k^{x,\mathrm{W}},p_k^{y,\mathrm{W}}]^T$ $[\mathrm{m}]$ and velocity $\bfv_k^\mathrm{W}=[v_k^{x,\mathrm{W}},v_k^{y,\mathrm{W}}]^T$ $[\mathrm{m}\,\mathrm{s}^{-1}]$ in a \textit{world} (W) frame aligned with the geographic north, i.e., $\bfx_k = [ (\bfp_k^\mathrm{W})^T, (\bfv_k^\mathrm{W})^T ]^T$.
The measurement $\bfz_k$ with $n_z=2$ is given by the barometric altimeter, which reads the vehicle altitude above the mean sea level (MSL) $\hslash_k^\mathrm{MSL}$ $[\mathrm{m}]$, and the odometer, which provides the vehicle velocity in the \textit{body} (B) frame $\bfv_k^\mathrm{B}=[v_k^{x,\mathrm{B}},v_k^{y,\mathrm{B}}]^T$ $[\mathrm{m}\,\mathrm{s}^{-1}]$. We assume that the heading of the vehicle is aligned with $\bfv_k^\mathrm{B}$; consequently, the W and B frames are rotated by the heading angle $\psi_k$ (angle between $\bfv_k^\mathrm{B}$ and $\bfv_k^\mathrm{W}$) and the respective direction cosine matrix (DCM) is $\bfC_k=\left[\begin{smallmatrix}
	\cos(\psi_k) & -\sin(\psi_k) \\ \sin(\psi_k) & \cos(\psi_k)
\end{smallmatrix}\right]$. The model \eqref{eq:ass} thus reads
\begin{subequations}
\begin{align}
	\bfx_{k+1} &= \left[\begin{matrix}
		\bfp_k^{\mathrm{W}} \\
		\bfv_k^{\mathrm{W}} 
	\end{matrix}\right]=\left[\begin{matrix}
		1 & 0 & 1 & 0 \\
		0 & 1 & 0 & 1 \\
		0 & 0 & 1 & 0 \\
		0 & 0 & 0 & 1
	\end{matrix}\right] \bfx_k + \bfw_k,\label{eq:4Dmodeldyn}\\
	\bfz_k&=\left[\begin{matrix}
		\hslash_k^{\mathrm{MSL}} \\
		\bfv_k^{\mathrm{B}} 
	\end{matrix}\right]=\left[\begin{matrix}
		\mathrm{terMap}(\bfp_k^{\mathrm{W}}) \\
		\bfC_k\bfv_k^{\mathrm{W}} 
	\end{matrix}\right]+\bfv_k,\label{eq:4Dmodel},
\end{align}
\end{subequations}
where $\mathrm{terMap}(\cdot)$ is the Digital Terrain Model of the Czech Republic of the 5th generation (DMR 5G) provided by the Czech State Administration of Land Surveying and Cadastre under license CC BY 4.0.
The ``true'' state was measured by GNSS receiver EVK-7 u-blox 7 Evaluation Kit for the \textit{reference purposes}. 
The values of $\hslash_k^{\mathrm{MSL}}$ were measured by MicroStrain 3DM-CV5-AHRS, and $\bfv_k^{\mathrm{B}}$ were simulated by perturbing the GNSS data with noise.

In this paper, both covariance matrices $\mathbf{R}_k$ and $\mathbf{Q}_k$ of the measurement noise $\mathbf{v}_k$ and dynamics noise $\mathbf{w}_k$, respectively, are assumed to be diagonal. While the diagonality of $\mathbf{R}_k$ is standard, the diagonality of $\mathbf{Q}_k$ is required by the proposed method—a limitation we aim to address in future work. 
Conditionally on $\bfx_k$, we thus have
\begin{subequations}
    \begin{align}
    p( \bfx_{k+1} | \bfx_k ) &= p\!\left( p_{k+1}^{x,\mathrm{W}}, v_{k+1}^{x,\mathrm{W}} \Big| p_k^{x,\mathrm{W}}, p_{k}^{x,\mathrm{W}} \right) \times \notag\\
    & \hspace{1cm} p\!\left( p_{k+1}^{y,\mathrm{W}}, v_{k+1}^{y,\mathrm{W}} \Big| p_k^{y,\mathrm{W}}, p_{k}^{y,\mathrm{W}} \right) \, , \label{eq:cond_indep-state} \\
    p( \bfz_k | \bfx_k ) &= p\!\left( \hslash_k^{\mathrm{MSL}} \Big| \bfp_k^{\mathrm{W}} \right) \cdot p\!\left( \bfv_k^{\mathrm{B}} \big| \bfv_k^{\mathrm{W}} \right)  \, ,
    \label{eq:cond_indep-meas}
\end{align}
\end{subequations}
where terms on each right-hand side involve disjoint state variables from the given $\bfx_k$.
Such independent structure is essential for the efficiency of the proposed method.

\subsection{Measurement update}
The likelihood function $p( \bfz_k | \,\cdot\, )$~\eqref{eq:cond_indep-meas} in tensor notation is
\begin{align}
	P_{\bfz_k|\bfx_k} = P_{\bfz_k^{(1)}|\bfp^\mathrm{W}_k} \odot P_{\bfz_k^{(2:3)}|\bfv^\mathrm{W}_k}  \, .
\end{align}
Each of the likelihood tensors $ P_{\bfz_k^{(1)}|\bfp^\mathrm{W}_k}$ and $ P_{\bfz_k^{(2:3)}|\bfv^\mathrm{W}_k}$ varies only along two state dimensions.
That is, the unique likelihood values for each likelihood tensor are given by matrices
\begin{subequations}
    \begin{align}
		\bfP_{\bfz_k^{(1)}|\bfp^\mathrm{W}_k}^{(i_1, i_2)} &= p_{\bfv_k}^{(1)} \left( \textbf{z}^{(1)}_k - \mathrm{terMap}\big( 
            \bfxi_{k, (i_1, i_2, 1, 1) }^{(1:2)}
        \big)\right)  \label{eq:auxone}\\
		 \bfP_{\bfz_k^{(2:3)}|\bfv^\mathrm{W}_k}^{(i_3, i_4)} &= p_{\bfv_k}^{(2:3)} \left(\textbf{z}^{(2:3)}_k - \bfC_k \cdot 
            \bfxi_{k, (1, 1, i_3, i_4)}^{(3:4)}
         \right) \, , \label{eq:auxtwo}
\end{align}
\end{subequations}
where the $p_{\bfv_k}^{(1)}$ is the one-dimensional noise PDF for the first dimension of the measurement and the $p_{\bfv_k}^{(2:3)}$ is the two-dimensional noise PDF for second and third measurement dimension.
The tensors $ P_{\bfz_k^{(1)}|\bfp^\mathrm{W}_k}$ and $ P_{\bfz_k^{(2:3)}|\bfv^\mathrm{W}_k}$ that have four independent indices can be constructed from $\bfP_{\bfz_k^{(1)}|\bfp^\mathrm{W}_k}$~\eqref{eq:auxone} and $\bfP_{\bfz_k^{(2:3)}|\bfv^\mathrm{W}_k}$~\eqref{eq:auxtwo} respectively, using~\eqref{eq:svdForCPD}.

Likelihoods and posterior weights are now both represented in CPD format. Therefore, the measurement update is performed using \eqref{eq:measUpdateTen} and \eqref{eq:hadCPD} iteratively with
\begin{align}
	P_{k|k} = P_{\bfz_k^{(1)}|\bfp^w_k} \odot P_{\bfz_k^{(2:3)}|\bfv^w_k} \odot P_{k|k-1} \, . \label{eq:posterior-tensor}
\end{align}
Since the resulting rank is the product of the ranks of all participating tensors, the resulting CPD tensor must be rank-reduced using the \textit{cp\_als} routine from the toolbox.

\subsection{Time-update}	
The proposed method adopts the Lagrangian form of the time update~\cite{MaDuSt:25}, consisting of two steps: advection and diffusion. In the remainder of this section, these steps are adapted to the CPD format by exploiting the independence structure in the dynamics equations~\eqref{eq:4Dmodeldyn}, particularly~\eqref{eq:cond_indep-state}.
		
\subsubsection{Advection Solution}
To solve the advection the grid points could be simply flown according to \eqref{eq:flow}.
However, this would result in a grid that is not aligned with the state-space axes at time $k+1$, rendering further computations inefficient and impractical.
This issue can be solved by
\begin{itemize}
    \item designing an axes-aligned predictive grid 
    \begin{align}
        \bfXi_{k+1} &= \Xi_{k+1}^{1} \times \cdots \times \Xi_{k+1}^{4} \, , \label{eq:axis-aligned-k+1}
    \end{align}
    covering the first two moments given by the Kalman filter (KF) prediction that is readily available for the considered  model with linear dynamics,
    \item transforming the grid $\bfXi_{k+1}$ back in time to yield\footnote{
Note that the grid $\widetilde{\bfXi}_k$~\eqref{eq:zmrsena-cela} cannot be expressed as a Cartesian product, i.e., it cannot be stored by dimensions. However, this is not an issue, as it does not have to be constructed in its entirety in practice; please see further.
    }
    \begin{align}
        \widetilde{\bfXi}_k \coloneqq \bfF_k^{-1}\bfXi_{k+1} \triangleq \{ \bfF_k^{-1} \bfxi_{k+1,j}; \ \bfxi_{k+1,j} \in \bfXi_{k+1} \} 
        \label{eq:zmrsena-cela}
    \end{align}
    so that the flow~\eqref{eq:flow} holds, and
    \item interpolating the weights $P_{k|k}$ from the grid $\bfXi_k$ to weights $\widetilde{P}_{k|k}$ on the interpolation grid $\widetilde{\bfXi}_k $ which we denote by
    \begin{align}
    	\bfXi_k,P_{k|k} \xrightarrow{\text{interpolation}} \widetilde{\bfXi}_k , \widetilde{P}_{k|k} \, , \label{eq:interpGood}
    \end{align}
    in a way suitable for the CPD format.
\end{itemize}
The first two points are straightforward while the last one is described in detail further.

\begin{figure}[]
	\centering
	\includegraphics[width=0.68\linewidth]{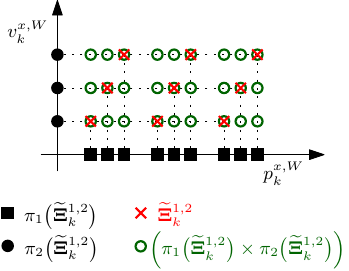}\vspace*{-2mm}
	\caption{Illustration of different grids and their projections for $x$ world direction at time step $k$.}
	\label{fig:gridProjection}\vspace*{-4mm}
\end{figure}

Noting that the weights we are interpolating from are stored in the CPD format as
\begin{align}
	{P}_{k|k} \approx \sum_{r=1}^{R} \lambda^{(r)} \mathbf{p}_1^{(:,r)} \circ \mathbf{p}_2^{(:,r)} \circ \mathbf{p}_{3}^{(:,r)} \circ \mathbf{p}_{4}^{(:,r)} \, , \label{eq:Pkk-CPD-pred-interpolaci}
\end{align}
we can interpolate the loading vectors for each rank along each axis (mode) separately. 
To represent the interpolated weights $\widetilde{P}_{k|k}$ in the CPD format without the need to work with the full tensor, we can use the independency~\eqref{eq:cond_indep-state}.
It follows that two tensors of order two, i.e., matrices in our case, must suffice to describe $\widetilde{P}_{k|k}$, and thus the interpolation in $x$ and $y$ world directions can be dealt with individually.
For advection, the key property is that the dynamics matrix is (after proper re-ordering of state elements) block diagonal
\begin{align}
   \begin{bmatrix}
       p_{k+1}^{x,\mathrm{W}} \\ v_{k+1}^{x,\mathrm{W}} \\ p_{k+1}^{y,\mathrm{W}} \\ v_{k+1}^{y,\mathrm{W}}    \end{bmatrix} = \begin{bmatrix}
           1 & 1 & 0 & 0 \\
           0 & 1 & 0 & 0 \\
           0 & 0 & 1 & 1 \\
           0 & 0 & 0 & 1 
       \end{bmatrix} \begin{bmatrix}
       p_{k}^{x,\mathrm{W}} \\ v_{k}^{x,\mathrm{W}} \\ p_{k}^{y,\mathrm{W}} \\ v_{k}^{y,\mathrm{W}}    \end{bmatrix}. 
\end{align}
To solve the interpolation in the $x$ world direction\footnote{
    The interpolation in the $y$ world direction is analogous.
}, first consider the two\discretionary{-}{-}{-}dimensional part of the interpolation grid $\widetilde{\bfXi}_k$~\eqref{eq:zmrsena-cela} for $p_k^{x,\mathrm{W}}$ and $v_k^{x,\mathrm{W}}$ only, i.e., after the appropriate re-ordering,
\begin{align}
	\widetilde{\bfXi}_k^{1,2} \coloneqq \begin{bmatrix}
	    1 & 1 \\ 0 & 1
	\end{bmatrix}^{-1} \left( {\Xi}_{k+1}^{1} \times  {\Xi}_{k+1}^{2} \right) \, , \label{eq:fhdjsak}
\end{align}
which is not axes-aligned as illustrated in Fig.~\ref{fig:gridProjection} (\GrCross{}).
Notice that the set $\pi_1 \big( \widetilde{\bfXi}_k^{1,2} \big) $ (\GrSquare{}) of projected points\footnote{
    The projection $\pi_{j}$ onto the $j$-th axis is defined as
    $\pi_{j} ( \bfXi ) = \cup_{\bfxi \in \bfXi} \, \{ \mathbf{e}_j^T \bfxi \} $, where 
    $\mathbf{e}_j = [0, \dots, 0,1,0, \dots, 0]^T$ has $1$ on the $j$-th entry. 
} on the first axis contain $\big| \pi_1 \big( \widetilde{\bfXi}_k^{1,2} \big) \big| = N_1 \cdot N_2$ number of position coordinates, while the set $\pi_2 \big(  \widetilde{\bfXi}_k^{1,2} \big)$ (\GrBullet{}) contains only $N_2$ velocity coordinates since $\left[\begin{smallmatrix}1&1\\0&1\end{smallmatrix}\right]^{-1} = \left[\begin{smallmatrix}1&-1\\0&1\end{smallmatrix}\right]$ in~\eqref{eq:fhdjsak} contains zero.
The loading vectors $\bfp_1^{(:,r)}$, $\bfp_2^{(:,r)}$~\eqref{eq:Pkk-CPD-pred-interpolaci} can now be interpolated from the old grid to new interpolating loading vectors on the interpolation grids as
\begin{subequations}
\begin{align}
	\Xi_k^{1},\ \mathbf{p}_1^{(:,r)} &\xrightarrow{\text{interpolate}} \pi_1 \big( \widetilde{\bfXi}_k^{1,2} \big), \ \widetilde{\mathbf{p}}_1^{(:,r)}, \quad \forall r \,, \\
	\Xi_k^{2},\ \mathbf{p}_2^{(:,r)} &\xrightarrow{\text{interpolate}} \pi_2 \big( \widetilde{\bfXi}_k^{1,2} \big), \ \widetilde{\mathbf{p}}_2^{(:,r)}, \quad \forall r \,,
\end{align}
\end{subequations}
which both are simple one-dimensional interpolations.
Using $\widetilde{\mathbf{p}}_1^{(:,r)}$ and $\widetilde{\mathbf{p}}_2^{(:,r)}$ directly as loading vectors, one could potentially construct a large tensor $\widetilde{\mathbf{p}}_1^{(:,r)} \circ \widetilde{\mathbf{p}}_2^{(:,r)} \in \mathbb{R}^{ (N_1\cdot N_2) \times N_2 }$ on the grid $ \pi_1 \big( \widetilde{\bfXi}_k^{1,2} \big) \times \pi_2 \big( \widetilde{\bfXi}_k^{1,2} \big)$ as illustrated on Fig~\ref{fig:gridProjection} (\GrCircle{}).
However, to solve the advection in a Lagrangian way, we are interested only in those entries in $\widetilde{\mathbf{p}}_1^{(:,r)} \circ \widetilde{\mathbf{p}}_2^{(:,r)}$ $\forall r$ that correspond to the desired subgrid $\widetilde{\bfXi}_k^{1,2} \subset \left( \pi_1 \big( \widetilde{\bfXi}_k^{1,2} \big) \times \pi_2 \big( \widetilde{\bfXi}_k^{1,2} \big) \right)$, see Fig.~\ref{fig:gridProjection} (\GrCross{}), (\GrCircle{}).
In practice, the corresponding matrices (on the mentioned subgrid) denoted as $\widetilde{\mathbf{P}}_{1,2}^{r} \in \mathbb{R}^{N_1 \times N_2}$ can be constructed $\forall r$ by back-projecting the points $\pi_1 \big( \widetilde{\bfXi}_k^{1,2} \big)$ into the two dimensions.
The back-projection can be realized by a mapping $\mu$ of one\discretionary{-}{-}{-}dimensional indices of points in $\pi_1 \big( \widetilde{\bfXi}_k^{1,2} \big)$ to two-dimensional indices of points in $\widetilde{\bfXi}_k^{1,2}$: for the $i$-th point of $\pi_1 \big( \widetilde{\bfXi}_k^{1,2} \big)$ there exist an index $(i_1,i_2) = \mu(i)$ such that 
\begin{align}
    \widetilde{\mathbf{P}}_{1,2}^{r,(i_1,i_2)} = \widetilde{\mathbf{p}}_1^{(i,r)} \cdot  \widetilde{\mathbf{p}}_2^{(i_2,r)} \, , \ \forall r \, . \label{eq:ooopop}
\end{align}
Unfortunately, the matrix $\widetilde{\mathbf{P}}_{1,2}^{r}$ regardless of $r$ cannot\footnote{
    The reason is that $i$ on the right-hand side of~\eqref{eq:ooopop} depends on $i_2$ via $i=\mu^{-1}(i_1,i_2)$, i.e., both product terms on the right-hand side of~\eqref{eq:ooopop} depends on $i_2$.
} be written as an outer product of some vectors $\bfa\in \mathbb{R}^{N_1}$ and $\bfb\in \mathbb{R}^{N_2}$ constructed from the loading vectors $\widetilde{\mathbf{p}}_1^{(i,r)} $ and $ \widetilde{\mathbf{p}}_2^{(i_2,r)}$ by omitting some of its entries, respectively.
However, the SVD~\eqref{eq:svd} is readily available for the matrices $\widetilde{\mathbf{P}}_{1,2}^{r}$ $\forall r$ to yield the CPD format
\begin{align}
    \widetilde{\mathbf{P}}_{1,2}^{r} = \sum_{r_x = 1}^{R_x^r} \mathbf{S}_{x}^{r,(r_x,r_x)} \cdot \mathbf{U}_{x}^{r,(:,r_x)} \circ \mathbf{V}_{x}^{r,(:,r_x)},
\end{align}
where $R_{x}^r$ is the rank of the SVD and $\mathbf{U}_{x} \in \mathbb{R}^{N_1 \times R_x}$ and $\mathbf{V}_{x} \in \mathbb{R}^{N_2 \times R_x}$ are the SVD factor matrices.

Repeating the above process for the remaining direction $y$ yields the matrices $\widetilde{\mathbf{P}}_{3,4}^{r}$ $\forall r$ with the decompositions
\begin{align}
    \widetilde{\mathbf{P}}_{3,4}^{r} = \sum_{r_y = 1}^{R_y^r} \mathbf{S}_{y}^{r,(r_y,r_y)} \cdot \mathbf{U}_{y}^{r,(:,r_y)} \circ \mathbf{V}_{y}^{r,(:,r_y)},
\end{align}
where $R_{y}^r$ is the rank of the SVD and $\mathbf{U}_{y} \in \mathbb{R}^{N_1 \times R_y}$ and $\mathbf{V}_{y} \in \mathbb{R}^{N_2 \times R_y}$ are the SVD factor matrices.
The desired tensor $\widetilde{P}_{k|k}$ on the entire 4-dimensional grid $\widetilde{\bfXi}_k$ in the CPD format can be constructed from the above decompositions using~\eqref{eq:svdForCPD}. 
Keeping the ordering of state dimensions as above, we have
\begin{align}
    \widetilde{P}_{k|k} &= \sum_{r=1}^{R} \sum_{r_x = 1}^{R_x^r} \sum_{r_y = 1}^{R_y^r} \lambda^{r} \cdot \mathbf{S}_{x}^{r,(r_x,r_x)} \cdot \mathbf{S}_{y}^{r,(r_y,r_y)} \ \times \notag\\
    & \hspace{1cm} \mathbf{U}_{x}^{r,(:,r_x)} \circ \mathbf{V}_{x}^{r,(:,r_x)} \circ \mathbf{U}_{y}^{r,(:,r_y)} \circ \mathbf{V}_{y}^{r,(:,r_y)} \nonumber \\
    &= \sum_{\rho = 1}^{\mathcal{R}} \widetilde{\lambda}^{(\rho)} \cdot \widetilde{\mathbf{q}}_1^{(:, \rho )} \circ \widetilde{\mathbf{q}}_2^{(:, \rho )} \circ \widetilde{\mathbf{q}}_3^{(:, \rho )} \circ \widetilde{\mathbf{q}}_4^{(:, \rho )} \, ,
\end{align}
where the the indices $r, r_x, r_y$ were mapped onto $\rho$ and
\begin{subequations}
\begin{align}
    \widetilde{\lambda}^{( r, r_x, r_y )} &= \lambda^{r} \cdot \mathbf{S}_{x}^{r,(r_x,r_x)} \cdot \mathbf{S}_{y}^{r,(r_y,r_y)} \, , \\
    \widetilde{\mathbf{q}}_1^{(:, (r,r_x,r_y) )} &= \mathbf{U}_{x}^{r,(:,r_x)} \, , \\
    \widetilde{\mathbf{q}}_2^{(:, (r,r_x,r_y) )} &= \mathbf{V}_{x}^{r,(:,r_x)} \, , \\
    \widetilde{\mathbf{q}}_3^{(:, (r,r_x,r_y) )} &= \mathbf{U}_{y}^{r,(:,r_y)} \, , \\
    \widetilde{\mathbf{q}}_4^{(:, (r,r_x,r_y) )} &= \mathbf{V}_{y}^{r,(:,r_y)} \, .
\end{align}
\end{subequations}
The resulting tensor $\widetilde{P}_{k|k}$ has rank $\mathcal{R} = \sum_{r=1}^{R} R_x^r R_y^r $, where $R$ is the rank of the original posterior $P_{k|k}$~\eqref{eq:Pkk-CPD-pred-interpolaci}. 
Since advection conserves the weights, it follows that the solution to the advection is given by the weights tensor 
\begin{align}
  P_{k+1|k}^{\text{adv}} = \widetilde{P}_{k|k}
\end{align}
on the grid $\bfXi_{k+1}$~\eqref{eq:axis-aligned-k+1}, which is axes aligned.

\emph{Remark:} It can be observed that the CPD rank grows rapidly during the advection process. Therefore, rank reduction is required at the end of advection, which is performed using the \textit{cp\_als} routine from the Tensor Toolbox. The authors also experimented with deflating the CPD weights after each estimation step for every rank $R$, which improved computational efficiency but did not yield sufficiently accurate estimates.

\subsubsection{Diffusion Solution}
Recalling that $\bfQ$ is assumed to be diagonal, the solution to the diffusion can be done by simple convolution of 1D Gaussian kernels (given by the dynamics noise) for each diagonal element of $\bfQ$ with the appropriate loading vectors. Therefore the loading vectors of diffusion solution $\mathbf{p}_j^{\text{dif},(:,r)}$ are calculated from the advection solution loading vectors $\mathbf{p}_j^{\text{adv},(:,r)}$ as\footnote{Convolution theorem would not lead to a significantly lower computational complexity as the convolution is only running on $N_j$ points in one dimension.} 
\begin{align}
   \mathbf{p}_j^{\text{dif},(:,r)} =  \mathbf{p}_j^{\text{adv},(:,r)} * W_{k,j},\ \forall r
\end{align}
where $W_j \in \mathbb{R}^{N_j}$ defined as
\begin{align}
    W_{k,j}^{(i_j)} = \mathcal{N} \big( \, (i_j-\bar{i}_j) \cdot \bfDelta_{k+1}^{(j)} ; \ 0, \bfQ^{(j,j)} \big)\, ,  \ \forall r
\end{align}
are the weights of PMD of Gaussian kernel for $j$-th state dimension given by the dynamics noise, where $\bar{i}_j = \lceil N_j /2 \rceil$ is the index of the middle grid point, c.f.,~$W_k$ in~\eqref{eq:lgbf-convolution}.

\section{Verification on Terrain-Aided Navigation}
This section presents results for the terrain-aided navigation scenario introduced earlier. A total of 20 Monte Carlo simulations were run, comparing the following filters:

\begin{itemize}
    \item Lagrangian GbF (LGbF) with $N = 51 \times 51 \times 41 \times 41 \approx 4{,}400{,}000$ \cite{MaDuSt:25},
    \item LGbF with spectral differentiation (LGbFs)  with the same $N$ \cite{MaDuGoGe:25},
    \item Bootstrap PF (PFb), with the same number of particles as points in the GbFs \cite{DoFrGo-book:01},
    \item UKF \cite{JuUhl:04},
    \item Proposed CPD-based LGbF (LGbF CPD) with $N_j = 101$, that is $N \approx 100{,}000{,}000$.
\end{itemize}

Note that the standard GbF is not part of the repository and and it would take hours to days for one step to be computed.
Also note that the Rao-Blackwellized PF available in the repository is not included, as it was unstable for this setup with forced diagonal process noise covariance matrix.

The filters are evaluated using the root mean square error (RMSE) for both position and velocity, along with the average computational time per time step, measured on a MacBook Air M1.
The results are presented in Table~\ref{tab:res}.
While the UKF is the fastest, it fails to accurately estimate either the position or the velocity.
The proposed method achieves the highest position estimation accuracy while also maintaining high velocity estimation accuracy.
Importantly, it also exhibits the lowest computational complexity, making it one of the least computationally demanding global methods developed to date. Moreover, it can be executed at least ten times per second, enabling real-time online estimation.

\emph{Remark:} Readers are encouraged to experiment with the published code; however, it should be noted that the proposed method may suffer from stability issues and negative PMD weights when used with a different set of user-defined parameters than those selected for this verification. Addressing this limitation remains an open topic for future research.

\begin{table}[h!]
\centering
\caption{Position and velocity RMSE, and average computational time per step.}
\begin{tabular}{lccc}
\toprule
Method & RMSE [$\mathrm{m}$] & RMSE [$\mathrm{m} \mathrm{s}^{-1} $] & Time [$\mathrm{s}$] \\
 & Position & Velocity & \\
\midrule
LGbF         & 15.23
   & 0.76
 & 3.88     \\
Spect LGbF   & 17.00
   & 0.84
 & 0.97     \\
PF bootstrap & 14.76
   & 0.82
 & 0.68     \\
UKF          & 290.97
    & 4.78
  & 0.0008  \\
LGbF CPD (Proposed)     & 14.13
   & 0.80
 & 0.06   \\
\bottomrule
\end{tabular}\label{tab:res}
\end{table}

\section{Summary and Future Work}

In this paper, we introduced a tensor decomposition-based grid estimation method that dramatically extends the applicability of grid-based filters. By leveraging the structure of the nearly constant velocity model and the terrain-aided navigation measurement equation, the method achieves remarkable computational efficiency—yet it remains general and can be adapted to arbitrary models with invertible dynamics, with complexity and accuracy dictated by the model structure. Most notably, the proposed approach scales linearly with the state dimension, breaking the long-standing curse of dimensionality that has limited grid-based filtering to low-dimensional problems. This advancement opens the door to applying grid-based filters in previously intractable scenarios enabling a new class of high-accuracy solutions across a wide range of applications.

In future work, we aim to develop an algorithm that eliminates the requirement for a diagonalized state noise covariance matrix, enabling greater flexibility and applicability in more general settings. We also plan to remove the assumption of linear dynamics, allowing for the estimation of models with arbitrary invertible dynamics. Furthermore, we intend to demonstrate the proposed method on higher-dimensional estimation problems to rigorously validate its linear computational scaling with respect to state dimension. Finally, there remain unresolved issues related to negative PMD weights and stability when user-defined parameters are not carefully tuned. Addressing these challenges will also be an important focus of future research.

\bibliographystyle{IEEEtran}
\bibliography{literatura}
\end{document}